\documentclass[twocolumn,twoside]{ncnsd}
  \usepackage{graphicx,psfrag}
  \def\BibTeX{{\rm B\kern-.05em{\sc i\kern-.025em b}\kern-.08em
      T\kern-.1667em\lower.7ex\hbox{E}\kern-.125emX}}

\begin{document}

\title{Residence Time Distribution of Sand Grains in the 1-Dimensional Abelian 
Sandpile Model}
%\author{Punyabrata Pradhan^{$\dagger$} \and Apoorva Nagar^{$\ddagger$}}
%\author{ Punyabrata Pradhan\thanks{E-mail: pradhan@theory.theory.tifr.res.in} ~and Apoorva Nagar\thanks{E-mail: apoorva@theory.theory.tifr.res.in}\\
%\\Department of Theoretical Physics,\\Tata Institute of Fundamental Research,\\ Mumbai, India 400005. }

%\maketitle
  \author{Punyabrata Pradhan and Apoorva Nagar
  \thanks{P. Pradhan (corresponding author)  and A. Nagar are with the Department of
  Theoretical Physics, Tata Institute of Fundamental Research, Homi Bhabha road,
  Mumbai - 400005, India. E-mail: pradhan@theory.tifr.res.in}}
  \markboth{NATIONAL CONFERENCE ON NONLINEAR SYSTEMS \& DYNAMICS} 
  {INDIAN INSTITUTE OF TECHNOLOGY, KHARAGPUR 721302, DECEMBER 28-30,
  2003}

\maketitle

%\begin{center}
%{Department of Theoretical Physics,\\Tata Institute of Fundamental Research,\\ Mumbai, India 400005}
%\end{center}

%\maketitle

%\begin{center}
%E-mail: ^{$\dagger$}pradhan@theory.theory.tifr.res.in, ^{$\ddagger$}apoorva@theory.theory.tifr.res.in
%\end{center}

%\begin{center}
%\mbox{~~~~~~~~} ^{$\ddagger$}apoorva@theory.theory.tifr.res.in
%\end{center}

%\begin{center}
%\bf{ABSTRACT}
%\end{center}
\begin{abstract}
We study the probability distribution of residence time, $T$, of the sand grains in the one dimensional abelian sandpile model on a lattice of $L$ sites, for $T<<L^2$ and $T>>L^2$. The distribution function decays as $\exp(-\frac{K_LT}{L^2})$. We numerically calculate the coefficient $K_L$ for the value of $L$ upto $150$ . Interestingly the distribution function has a scaling form $\frac{1}{L^a}f(\frac{T}{L^b})$ with $a \neq b$ for large $L$. 
\end{abstract}

%We show that even if $a$ and $b$ are not equal, it's consistent with the normalization of the distribution function since the scaling function $f(x)$ diverges when $x \rightarrow 0$ and has a lower cutoff at $x=\frac{1}{L^2}$. 

%PACS classification codes: 02.50.-r; 05.40.fb; 66.10.cb

\section{Introduction}
There are many slowly driven extended dissipative systems in nature which organize themselves to a critical state. These are called self organized critical (SOC) systems. Such systems show long-ranged spatial and temporal correlations, i.e., power law behaviour in the correlation functions; for example, the frequency of earthquakes as a function of energy released, the mean square velocity difference as a function of distance in the turbulant fluids, height-height correlation in the case of surface roughening of growing interfaces ~\cite{sornette,bak,maya,chen}. The time series of electrical noise shows power law in the power spectrum ( ``$1/f$'' noise)~\cite{bak}. Sandpile models are the simplest models where important features of Self Organized Critical (SOC) states are present. Abelian sandpile model is a subclass of these. There are many papers ~\cite{maya,amaral,kim,vespig} concerned mainly in studying the distributions of avalanche sizes and avalanche durations and the corresponding critical exponents, but there is not much work in the literature where the distribution of times these grains spend inside the system has been studied theoretically. Although it has been studied in an experiment on ricepiles and in simulation in one dimension ~\cite{kim}. In this paper we study the residence time distribution of a marked sand grain, when the system is driven by adding sand grains at one end (say, left). Our model is BTW sandpile model ~\cite{tang} in 1-dimension which is simplest to study. We find out the scaling form to be $\frac{1}{L^3}f(\frac{T}{L^2})$ for large $L$, where $T$ is the residence time and $L$ is the size of the system. This scaling function $f(x)$, although it diverges as $x \rightarrow 0$, is consistent with the normalization condition, since lower limit of $x$ has a cutoff at $x=\frac{1}{L^2}$. The residence time distribution decays exponentially for large value of $T$. We determine numerically the coefficient, $K_L$, of $T/{L^2}$ in the exponential decay for some finite sizes of the 1-diemnsional lattice. 
\bigskip

\section{Model}
We take an one dimensional lattice of $L$ sites where sites are indexed by $i$ with $1 \leq i \leq L$. The configuration of the sandpile is specified by height variable (number of sand grains) $z_i$ at any site $i$. The dynamics is as follows. Whenever the height is higher than some critical value $z_c$, the site topples and two sand grains go to the nearest neighbour sites, one goes to the left and the other goes to the right. Which one goes to the left is  chosen at random. Here we consider both $z_c=1$ and $z_c=2$. There is one layer of sand grains for $z_c=1$. For $z_c=2$, there are two layers of grains. When the height of the end site at $i=1$ or $i=L$ crosses the critical height, the site topples. One sand grain goes out of the system and other goes to the next neighbour site. No sand is added till the system comes to the stable configuration, i.e., there is no more topplings in the system. The time unit is measured as the duration between the successive addition of two sand grains. The residence time of a sand grain is defined as the time spent by that grain inside the system. $P(T|L)$ is the probability distribution function of the residence time $T$ where $L$ is the number of sites in the 1-dimensoinal lattice.  
\\
First we consider the model with critical height $z_c=1$ and start from the configuration with the height $z=0$ at all sites. Now we keep on adding sand grain at the leftmost site. After first addition of grain, there is no toppling. Then we add the next sand grain. Now there will be toppling, one sand grain will leave the system, other grain will move to the right and stays there until next grain comes to the site, and so on. So the evolution rule is totally deterministic. 
After some time ($\sim L^2$) the system falls into a cycle. This cycle, consisting of $L+1$ stable configurations, has at most one site with $z_i=0$ and other sites with $z_i=1$ (recurrent configuration space). In the steady state of the system, all the configurations are equally probable. %Whatever observable we want to calculate, we just consider the average of that observable with respect to the reccurent configuration space. 
\\
%The ``zero'' of the system is defined as the site $i$ having $z_i=0$. later on we will denote the position of ``zero'' $Y_0$. 
Of the $L+1$ recurrent configurations, $L$ configurations have a site with zero height and one configuration has all sites with height $1$. If there is a configuration which has $i$-th site with zero height, in the next time step the configuration changes to one which has $(i-1)$-th site with height zero (If $i=1$, in the next step all heights become $1$). At the next time step, the $(i-2)$-th site is with height zero. After some time there is 1st site with zero height. When next sand grain is added, all sites become with height $1$. In the next step, last site becomes with height zero and so on. Even though time evolution of the system is completely deterministic, movement of sand grains is stochastic. Whenever any site topples, one of the two sand grains goes to the left nearest neighbour and stops, and other grain moves to the right nearest neighbour. The probability that one particular grain of the two goes to the left and other goes to the right is $1/2$. If the height at the right site is zero, toppling stops, otherwise, there is another toppling. If any sand grain falls on the site with height zero, it stops moving and waits till the site with height zero is at the left of it. When the $L$-th site has the height zero, the sand grain starts moving again in the similar fashion. The waiting times are not same at every site and waiting times are correlated. 
\\

% EXTENSION at the end

%The case for $z_c=2$ is almost similar except for the fact that whenever $z>2$, the site topples and sand grains move to the left or right next neighbouring site with probability $\frac{1}{3}$. With probability $\frac{1}{3}$ grains stick to the site, i.e., do not move. We study the residence time distribution of a marked sand grain when the system is in the steady state.   

%In the steady state system passes with the time period $L+1$ through the recurrent configurations with only one zero at any site between $i=1$ to $i=L$. The 'zero' always moves one step to the left after each time step, vanishes at $i=1$ and again appears at $i=L$ perodically. In the steady state we add one marked sand at the left edge. Whenever the height at that sight crosses $z_c$, it goes to the neighbouring sites at random. If it  moves to the left, it stops. If the marked sand is at the left or right edge and moves further to the left or right respectively, it gets out of the system (i.e., absorbed). The residence time is the duration for which the marked sand spends inside the system in steady state. We are interested to find out the probability $P(T/L)$ that the marked sand gets absorbed at time $T$ where $L$ is the lattice size.
 
\section{Analytical Results}
First we shall discuss the case for $z_c=1$. When the residence time $T$ is very small compared to $L$, sand grains always get ejected from the left and do not go beyond the site $i=T$ (except for the case when $z_i=1$ at all sites at time $T$ and grain might be ejected from the right end). It's very unlikely that the marked sand grain meets the site with height zero  (the probability $\leq \frac{T}{(L+1)}$ which goes to zero for large value of $L$). In the limit of  $L$ large compared to $T$, the residence time distribution is well approximated by the distribution of the first passage (at $i=1$) time of a simple unbiased random walker with $2T$ steps random walk. So, for $T<<L$, We can exactly calculate the residence time distribution ~\cite{feller}
$$
P(T|L)=  \frac{2T!}{T!(T+1)!} 2^{-(2T+1)}
$$
where  $P(T|L)$ is the probability that the marked sand grain gets ejected from the system immediatly after time $T$. Using Sterling approximation to the above expression, we get the function $P(T|L)$  equals to $\frac{1}{2\sqrt{\pi}}\frac{1}{T^{3/2}}$ for $1<<T<<L$. Now we can extend our result whenever $T<<L^2$. The argument is as follows. After time steps $T \sim L^ \alpha$ where $\alpha<2$, the standard deviation of the position of the marked sand grain goes as $L^ {\alpha/2}$. So the probability that the marked grain will meet the site with height zero goes as $L^ {\alpha/2-1}$ which will tend to zero in the large $L$ limit. Therefore for any $1<<T<<L^2$, the limiting probability distribution $P(T|L)$ is proportional to $T^{-3/2}$, the first passage (at $i=1$) time distribution of a simple unbiased random walker with $2T$ steps random walk. For the large value of $T$ the distribution function decays exponentially like $\exp(-K_LT/L^2)$ where $K_L$ depends on lattice size. Later we will describe how to compute the coefficient $K_L$ for any finite $L$. 
\\

%EXTENSION at the end for writing below

%The case for $z_c=2$ is much similar to the previous case except that sand grain can stick to the site with probablity $1/3$ even though there is toppling and with probability $\frac{1}{3}$ it goes to the right or left. The residence time distribution $P(T|L)$, for $1<<T<<L$, is 
%\begin{equation}
%\sum_{T'=0}^{T} ^{(T+T')}C_{2T'}3^{-(T-T')}\frac{2T!}{T!(T+1)!} 3^{-(2T+1)} \mbox{~~}\sim \mbox{~~}\frac{1}{T^{3/2}} 
%\label{eqn1}
%\end{equation}
%Here also we can extend this result for $1<<T<<L^2$ just as we did in the case for $z_c=1$.

Even though we don't know the distribution function for all $T$, we can find out the first moment of the residence time distribution easily. We define mass of the sandpile as the total number of particles in the pile (i.e., $\sum_{i=1}^{i=L}z_i$). It's easy to see that mean residence time $\langle T \rangle =\langle \mbox{Total mass of the pile} \rangle$. To prove this, let us define an indicator funtion $\eta_{n,T}$ as given below.
$$
\eta_{n,t}=1 \mbox{ ~~if the sand grain, added at time $n$, is in the system }
$$
$$\mbox{at time $t$, otherwise $\eta_{n,t}=0$.~~~~~~~~~~~~~~~~~~~~~~~~~~~~~~~~~~~~~~~~~~~~~~~~~~~~}
$$

%$$ \mbox{otherwise $\eta_{n,t}=0$.}  
%$$

The mean residence time can be written as
$$ 
\langle T \rangle=\lim_{\mathcal{N},\mathcal{T} \rightarrow \infty} \frac{1}{\mathcal{N}} \sum_{n=1}^{\mathcal{N}} \sum_{t=1}^{\mathcal{T}} \eta_{n,t}=\frac{1}{\mathcal{N}} \sum_{t=1}^{\mathcal{T}} \sum_{n=1}^{\mathcal{N}} \eta_{n,t}
$$
$$
=\frac{1}{\mathcal{N}} \sum_{t=1}^{\mathcal{T}} \mbox{(Total mass of the pile at time $t$)}
$$
$$
= \langle \mbox{~Total mass of the pile~} \rangle \mbox{~~~~~~~~~~~~~~~~~~~~~~~}
$$
The average of the total mass in the pile is $[\frac{L}{L+1}(L-1)+\frac{1}{L+1}L]$ which goes as $L$ for large $L$. 
%Since the quantity $\sum_{n=1}^{N} \eta_{n,t}$ has lower bound $(L-1)$ and upper bound $L$ for any time $t$, $\frac{T}{N}.(L-1) \leq \langle T \rangle \leq \frac{T}{N}.L$. In the limit $T,N \rightarrow \infty$, $\frac{T}{N} \rightarrow 1$ and hence $\langle T \rangle \sim L$. 
\\
We find out the scaling form of the function $P(T|L)$ using the mean residence time and the previous limiting distribution. The scaling finction $f(x)$ is defined as 
$$
f(x)dx=\lim_{L\rightarrow \infty} L^{a-b} \mbox{Prob}(x L^b \leq T \leq (x+dx)L^b)
$$
So the form of the distribution function $P(T|L)$ is $\frac{1}{L^a}f(\frac{T}{L^b})$. Since $P(T|L)$ varies as $T^{-3/2}$ for $1<<T<<L$, $f(x)$ must goes as $x^{-3/2}$ for very small value of $x$ and therefore $\frac{1}{L^a}f(\frac{T}{L^b})$ goes as $L^{(3b/2-a)}T^{-3/2}$. As $P(T|L)$ is independent of $L$ for $1<<T<<L$, we get $a/b=3/2$. Even though the scaling function $f(x)$ is divergent as $x$ tends to zero, we can normalize $P(T|L)$ without any problem, since $x$ has lower cutoff $\frac{1}{L^b}$. As $f(x)$ goes as $x^{-3/2}$, the normalisation integral is 
$$
L^{b-a} \int_{1/L^b}^\infty f(x)dx \sim L^{3b/2-a}
$$
 which is independent of $L$ as $a/b=3/2$. The mean residemce time $\langle T \rangle$ is given by
$$\int _0^\infty \frac{1}{L^{3b/2}}f(T/L^b)TdT \sim L^{b/2} $$
%$$\mbox{(Since there is no divergence in the integrand, }
%$$
%$$ \mbox{~~~~~~~~~~~~~~~~~~~~~~~~~~~~~~we put lower limit in the integration zero.)}
%$$
As there is no divergence in the integrand, we can put the lower limit in the integration zero.
Since $\langle T \rangle \sim L$, we find $b=2$ and therefore can write the scaling form of the $P(T|L)$ as $T^{-3/2} \tilde f (\frac{T}{L^2})$ for large $L$. In Fig.1. probability distributions, $P(T|L)$, of residence time of sand grains are plotted for different lattice sizes $L=25,50,100,150$. In Fig.2. all the curves for $P(T|L)$ collapse to a single curve when we plot $L^3P(T|L)$ against $T/{L^2}$.

%Now we can extend our result whenever $T<<L^2$. The argument is as follows. After time steps $T \sim L^ \alpha$ where $\alpha<2$, the standard deviation of the position of the marked sand grain is less than some constnat times $\sqrt(L^ \alpha)$. The probability that the marked grain will meet the 'zero' is $\frac {\sqrt(L^ \alpha)} {L}$ which will tend to zero for $L>>1$. Therefore for any $T<<L^2$ and $L>>1$ the limiting probability distribution $P(T/L)$ is the first passage time distribution $T^{-3/2}$. 

\section{Computational Results}
Now we study the behaviour of the function $P(T|L)$ when $T>>L^2$. Since time evolution of states of the system is a Markov process, there is exponential decay of probabilities of states for large value of time, due to the presence of the absorbing boundaries. So the probability distribution $P(T|L)$ must decays as $\exp(-K_LT/{L^2})$ for very large $T$. For large $L$, the coefficient, $K_L$, tends to a constant $K$ and scaling function $f(x)$ must goes as $\exp(-Kx)$ for large $x$. We define $K$ as a limit given below,
$$
K=-\lim_{L \rightarrow \infty} \mbox{~}[ \mbox{~}\lim_{T \rightarrow \infty} \frac{L^2 \ln{P(T|L)}}{T} \mbox{~}]
$$
We numerically calculate values of $K_L$'s for various finite lattice sizes by defining transition probability matrices for the system going from one recurrent configuration to another. Now we need to distinguish all configurations with a specific site which has height zero. Configurations are distinguishable with respect to the position of the marked grain. Marked grain can be at any site except the site where height is zero. For simplicity, we construct the transition matrices for small value of $L$, say $L=4$. When the site with zero height is at the end, i.e., at $i=4$, three distinguishable configurations are represented respectively  as $|\ast 110 \rangle$, $|1\ast10\rangle$, $|11\ast 0\rangle$ where ``Star'' denotes the position of marked grain, ``$1$'' denotes the site with height one and ``$0$'' denotes the site with height zero. If we keep on adding sand at the left, the transitions will occur from one state to another. In this case transitions will be from
$$
\{1110\} \mbox{~~} \rightarrow \mbox{~~} \{1101\} \mbox{~~} \rightarrow \mbox{~~} \{1011\} \mbox{~~} \rightarrow \mbox{~~} \{0111\} \mbox{~~} 
$$
$$
\rightarrow \mbox{~~} \mbox{~}\{1111\} \rightarrow \mbox{~~}\mbox{~}\{1110\}
$$  
%$$
%\{\ast 110\}, \{1\ast10\}, \{11\ast 0\} \mbox{~~} \Rightarrow \mbox{~~} \{\ast 101\}, \mbox{~}\{1\ast01\}, \mbox{~}\{110\ast \}
%$$
%$$
% \Rightarrow \mbox{~~} \{\ast 0 11\}, \mbox{~}\{10\ast1\},\mbox{~}\{101\ast \} \mbox{~~} \Rightarrow \mbox{~~} \{0\ast 11\}, \mbox{~}\{01\ast1\},\mbox{~}\{011\ast \}
%$$
%$$\Rightarrow \mbox{~~} \{1\ast 11\}, \mbox{~}\{11\ast1\}, \mbox{~}\{111\ast \} \mbox{~~} \Rightarrow \mbox{~~} \{\ast 110\}, \mbox{~}\{1\ast10\}, \mbox{~}\{11\ast 0\} 
%$$
$$\mbox{ (same initial states after $L+1=5$ time steps)}.$$
Here marked grain can be at any one of the sites with height $1$. As mentioned before, we represent basis states for a particular configuration as $|j,Y_0\rangle $ where $j$ denotes the position of marked particle and $Y_0$ denotes the index of the site with height zero. Since marked grain cannot stay at the site with height zero, there are $L-1$ basis states for a particular configuration (i.e. for a fixed $Y_0$). $Y_0$ can take value from $0$ to $L$. $Y_0=0$ means that all sites are with height $1$. Whenever we add sand grain at the left end, there is transition from any of the basis states with some value of $Y_0$ to any of the basis states with $Y_0-1$  (i.e., $|j,Y_0 \rangle \rightarrow |i,Y_0-1 \rangle$). $i$ and $j$ denote the position of marked sand grain. Now we define elements of the transitoin matrix,${\mathcal{T}}_{Y_0}(i|j)$, as the transition probability from the $j$-th state to $i$-th state where $Y_0$ denotes that the transition occurs from the configuration with height zero at $Y_0$-th site to the configuration with height zero at $(Y_0-1)$-th site.  The transition matrices ${\mathcal{T}}_4$,  ${\mathcal{T}}_3$, ${\mathcal{T}}_2$, ${\mathcal{T}}_1$, ${\mathcal{T}}_0$ for $L=4$ are written below explicitly.       
%At each time step the transition between any two states across the arrow is allowed.  

%The Matrix $T_4$:
%\begin{equation}
%$$
\[ {\mathcal{T}}_4=
\left[ \begin{array}{ccc}
                      
       (\frac{1}{2})^{2}  & \frac{1}{2}       & 0           \\      
        (\frac{1}{2})^{3} & (\frac{1}{2})^{2} & \frac{1}{2} \\
        (\frac{1}{2})^{3} & (\frac{1}{2})^{2} & \frac{1}{2}
   \end{array} \right] \mbox{~~~~~~~~~~~~~~~~}
   {\mathcal{T}}_3=
\left[ \begin{array}{ccc}
                      
       (\frac{1}{2})^{2}  & \frac{1}{2}       & 0   \\      
        (\frac{1}{2})^{2} & \frac{1}{2} & 0 \\
        0 & 0 & 1
   \end{array} \right]    
\]
%$$
%\end{equation} 

\[ {\mathcal{T}}_2=
\left[ \begin{array}{ccc}
                      
       \frac{1}{2}  & 0  & 0  \\      
        0 & 1 & 0 \\
        0 & 0 & 1
   \end{array} \right] \mbox{~~~~~~~~~~~~~~~~~~~~~~~~}
  {\mathcal{T}}_1=
\left[ \begin{array}{ccc}
                      
       1  & 0  & 0           \\      
       0 & 1 & 0 \\
       0 & 0 & 1
   \end{array} \right] 
\]

\[
 {\mathcal{T}}_0=
\left[ \begin{array}{ccc}
                      
       \frac{1}{2}  & 0  & 0           \\      
       (\frac{1}{2})^{2} & \frac{1}{2} & 0 \\
       (\frac{1}{2})^{3} & (\frac{1}{2})^{2} & \frac{1}{2}
   \end{array} \right]    
\]

%The general structure of the transition matrices $\mathcal{T}_{Y_0}$ for any lattice size $L$ with ``zero'' at $i=Y_0$ is given below.

% \[ \mathcal{T}_{Y_0}=
%\left[ \begin{array}{cccccccccccc}
                      
%       (\frac{1}{2})^2  & \frac{1}{2}  & 0 & 0 & \cdot & 0 & 0  & \cdot & \cdot & \cdot & \cdot & 0 \\      
%       (\frac{1}{2})^{3} & (\frac{1}{2})^2 & \frac{1}{2} & 0 & . & 0 & 0 & \cdot & \cdot & \cdot & \cdot & 0 \\
%       . & . & . & \frac{1}{2} & 0 & 0   & 0 & \cdot & \cdot & \cdot & \cdot  & 0 \\
%       . & . & . & . & \frac{1}{2} & 0 & 0 & \cdot & \cdot & \cdot & \cdot & 0 \\
%       (\frac{1}{2})^{Y_0-1} & (\frac{1}{2})^{Y_0-2} & . & . & (\frac{1}{2})^2 &  \frac{1}{2}  & 0 & \cdot & \cdot & \cdot & \cdot & 0 \\
%       (\frac{1}{2})^{Y_0-1} & (\frac{1}{2})^{Y_0-2} & . & . & (\frac{1}{2})^2 &  \frac{1}{2} & 0 & \cdot & \cdot & \cdot & \cdot & 0 \\
%       0 & \cdot & \cdot & \cdot & \cdot & 0 & 1 & 0 & 0 & 0 & 0 & 0 \\
%       \cdot & \cdot & \cdot & \cdot & \cdot & \cdot & 0 & 1 & 0 & 0 & 0 & 0 \\
%       \cdot & \cdot & \cdot & \cdot & \cdot & \cdot & 0 & 0 & 1 & 0 & 0 & 0 \\
%       \cdot & \cdot & \cdot & \cdot & \cdot & \cdot & 0 & 0 & 0 & 1 & 0 & 0 \\ 
%       \cdot & \cdot & \cdot & \cdot & \cdot & \cdot & 0 & 0 & 0 & 0 & 1 & 0 \\
%       0 & \cdot & \cdot & \cdot & \cdot & 0 & 0 & 0 & 0 & 0 & 0 & 1 \\
%   \end{array} \right]    
%\]

The transition matrix ${\mathcal{T}}^{(L)}=\prod_{{Y_0}=0}^{L} {\mathcal{T}}_{Y_0}$ gives back the initial configuration after the period of time $L+1$. The superscript in ${\mathcal{T}}^{(L)}$ is denoting that the marked grain is added when the initial configuration is with $Y_0=L$. In general, if the grain is added with the initial configuration where position of the site with heigt zero is at $i=Y_0$, the transition matrix will be ${\mathcal{T}}^{(Y_0)}={\mathcal{T}}_{Y_0-1}{\mathcal{T}}_{Y_0-2}....{\mathcal{T}}_{L-1}{\mathcal{T}}_L{\mathcal{T}}_0..$ $..{\mathcal{T}}_{Y_0-1}{\mathcal{T}}_{Y_0}$. The different values of $Y_0$ in transition matrices correspod to the addition of sand grains to the configurations with different sites with height zero. To find out the coefficient  $K_L$ in the exponential, we diagonalize any matrix ${\mathcal{T}}^{(Y_0)}$, say ${\mathcal{T}}^{(L)}$, and take the largest eigenvalue, $\lambda_{max}$. We diagonalise any one matrix of the different transitoin matrices because the eigen values for the different transitoin matrices, ${\mathcal{T}}^{(Y_0)}$s (we get various ${\mathcal{T}}^{(Y_0)}$s just with cyclic permutation of matrices ${\mathcal{T}}_{0}$, ${\mathcal{T}}_{2}$, ${\mathcal{T}}_{3}$....${\mathcal{T}}_{L}$ etc.) are same. For large value of time $T$ ($>>L^2$), the residence time distribution decays exponentially like $(\lambda_{max})^{T/(L+1)}$ which equals to $\exp(\frac{T}{L+1}\ln\lambda_{max})$. The coefficient $K_L$ is then $-\frac{L^2}{L+1}\ln\lambda_{max}$. In the Fig.3. we have plotted $K_L$ against the number of sites in the 1-dimensional lattice, $L$. It shows that $K_L$ saturates to the value of $K$. So the coefficient is independent of lattice size $L$ for large value of $L$. 
\\
\\

\begin{figure}
 % \vspace{2.5cm}
  \centering
  \includegraphics[width=0.95\columnwidth,angle=0]{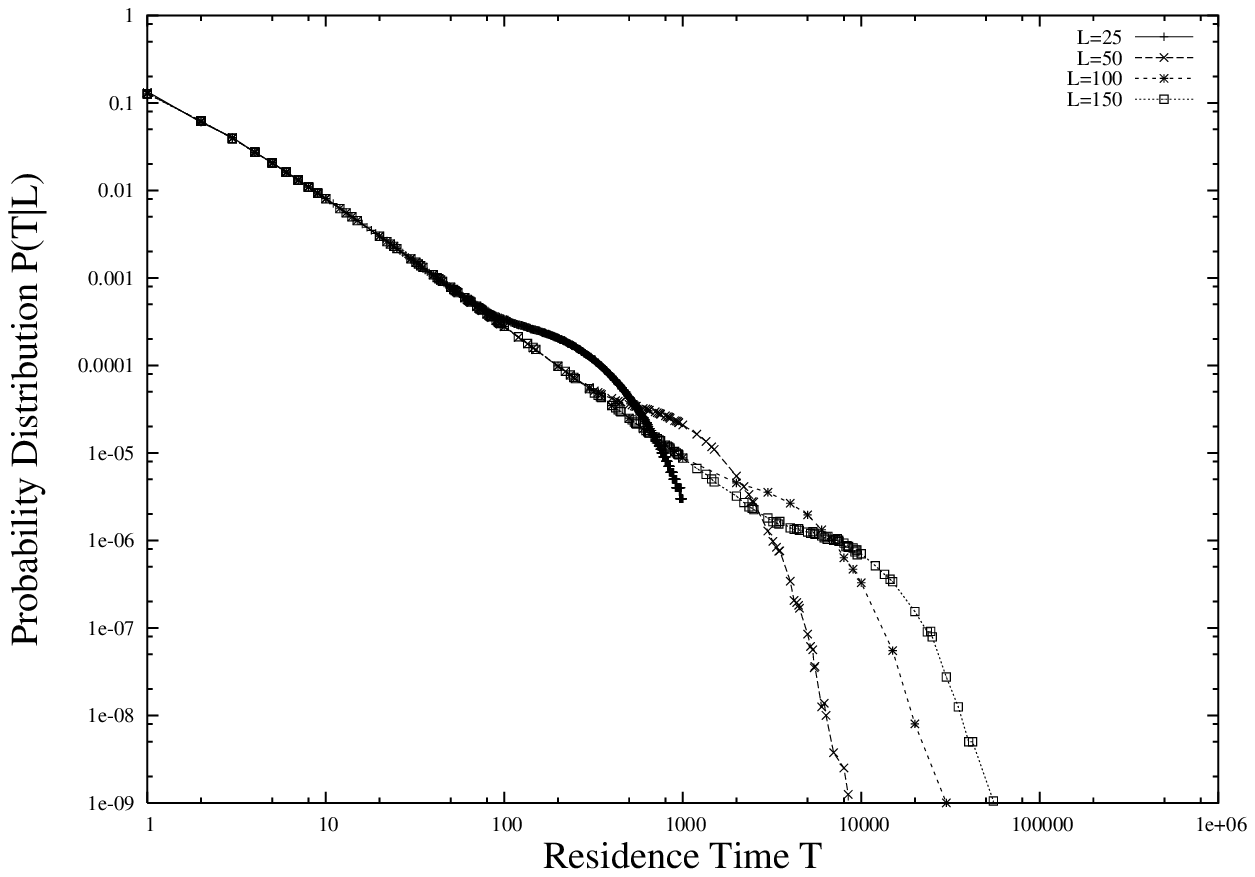}
  \caption{The probability distribution of residence time}
  \label{fig1label}
  \end{figure}

\begin{figure}
%\vspace{2.5cm}
  \centering
  \includegraphics[width=0.95\columnwidth,angle=0]{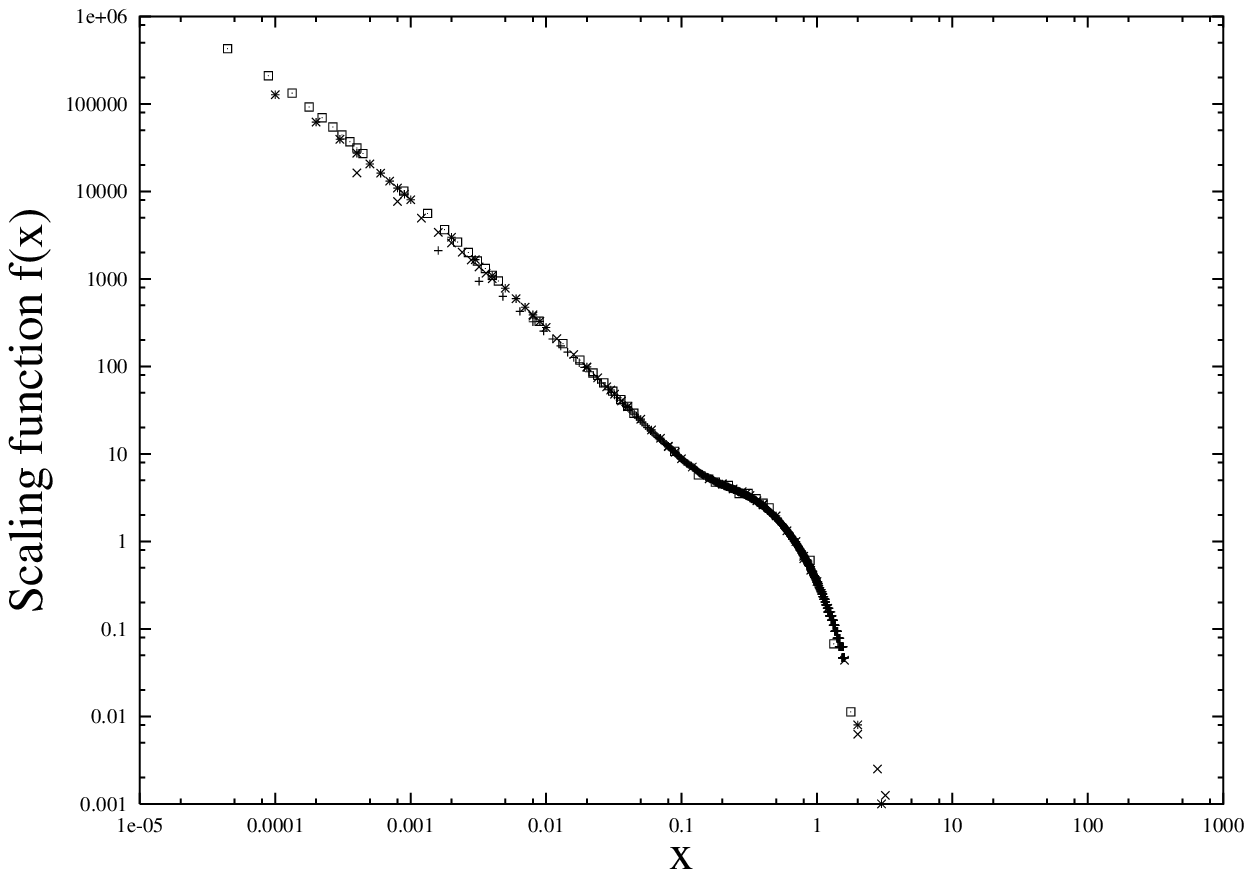}
  \caption{The plot of scaling function $f(x)$ showing collapse of all the curves for differnet lattice sizes to a single curve.}
  \label{fig1label}
  \end{figure}

%\vspace{2cm}
\begin{figure}
%\vspace{2.5cm}
  \centering
  \includegraphics[width=0.95\columnwidth,angle=0]{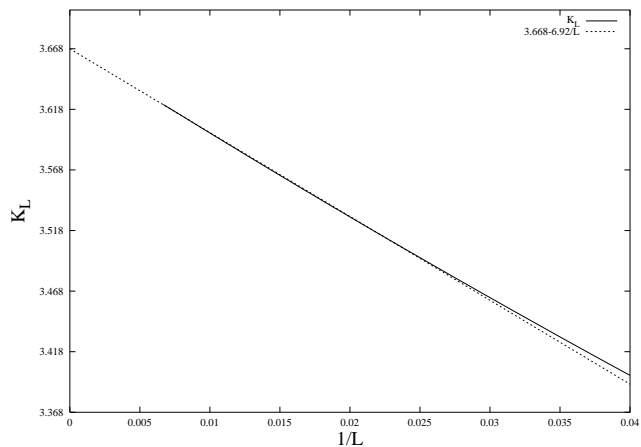}
  \caption{Plot of $K_L$ versus lattice size L}
  \label{fig1label}
  \end{figure}

%\newpage
%\section{Extension to the case for $z_c=2$}
We can easily extend all our previous result for $z_c=1$ to $z_c=2$. The toppling rules for $z_c=2$ are similar to the previous case $z_c=1$ except that sand grain can stay at the site even if there is a toppling. Whenever height at any site is three, the site gets unstable and topples. One sand grain of the three goes to right nearest neighbour site, one goes to the left nearest neighbour site and the other one stays put. It is decided at random which one goes to the right, which one goes to the left. The probablity that a particular grain of the three going to the right or to the left, or staying there is $1/3$. The height of any site is either $1$ or $2$, when the system is stable. In the steady state, at most one site has height $1$ and all the others have height $2$. All configurations in the recurrent configuration space are equally probable. The number of recurrent configuratins is same as before. The residence time distribution $P(T|L)$, for $1<<T<<L$, is 
\begin{equation}
\sum_{T'=0}^{T} {}^{(T+T')}C_{2T'}3^{-(T-T')}\frac{2T!}{T!(T+1)!} 3^{-(2T+1)} 
\mbox{~~}\sim \mbox{~~}\frac{1}{T^{3/2}} 
\label{eqn1}
\end{equation}
Here also we can extend this result for $1<<T<<L^2$, just as we did in the case for $z_c=1$. The probability distribution functin has the same scaling form as for $z_c=1$.

To calculate the coefficient $K_L$, we can construct the trasition probability matrix exactly in the similar way for the previous case. In this case the number of basis states is $L$ and $K_L$ can be determined as before.   
%\newpage
%Now we can find out all the transition matrices for the transition between two different configurations and multiply them to find out the maximum eigenvalue for any finite lattice size $L$. $K_L$ is related to the maximum eigenvalue as written before.   
 \\   

\section{Summary}
In summary, we have calculated probability distribution of the residence time $T$ in two limitting cases $T<<L^2$ and $T>>L^2$, have shown that the distribution function has a scaling form $\frac{1}{L^3}f(\frac{T}{L^2})$ and computed the coefficient of $T/{L^2}$, $K_L$, in the exponential decay of the distribution function for the values of $L$ upto $150$.
\\
\begin{center}
\bf{Acknowledgment}
\end{center}
We are deeply indebted to Prof. Deepak Dhar for guiding us at every step of this work.
\\

\bibliographystyle{ncnsd}

\begin{thebibliography}{1}


\bibitem{sornette}A. Sornette and D. Sornette, Self-organised criticality of earthquakes, Europhysics Letters \newblock{\bf9}, \newblock 197 (1989).
\bibitem{bak}P. Bak and K. Chen, The physics of fractals, Physica D \newblock{\bf38}, \newblock 5 (1989).
\bibitem{maya}M. Paczuski and Stefan Boettcher, Universality in Sandpiles, Interface depinning, and Earthquake models, Physical Review Letters \newblock{\bf77}, \newblock 111 (1996).
%\bibitem{bak}P. Bak and K. Chen, The physics of fractals, Physica D \newblock{\bf38}, \newblock 5 (1989).
\bibitem{chen}K. Chen, P. Bak and S. P. Obukhov, Self-organized criticality in a crack propagation model of earthquakes, Physical Review A \newblock{\bf43}, \newblock 625 (1991).
\bibitem{amaral}L. A. N. Amaral and K. Lauritsen, Self-organized criticality in a rice-pile model, Physical Review E \newblock{\bf54}, \newblock R4512 (1996).
\bibitem{kim}K. Christensen, A. Corral, V. Frette, J. Feder, and Torstein Jossang, Tracer dispersion in a Self-Organized Critical system, Physical Review Letters \newblock{\bf77}, \newblock 107 (1996).

\bibitem{vespig}A. Vespignani, Stefan Zapperi, How self-organized criticality works: A unified mean-field picture, Physical Review E \newblock{\bf57}, \newblock 6345(1998).
\bibitem{tang}P. Bak, C. Tang and K. Wisenfeld, Self-organized criticality: An explanation of the $1/f$ noise, Physical Review Letters, \newblock{\bf59}, 381, 1988. \\Self-organized criticality, Physical Review A,  \newblock{\bf38}, 364, 1988.
\bibitem{sen}P. Ruelle and S. Sen, Toppling distributions in one-dimensional Abelian sandpiles, J. Phys. A: Math. Gen., \newblock{\bf25}, L1257, 1992.
\bibitem{ali}A. A. Ali and D. Dhar, Breakdown of simple scaling in Abelian sandpile models in one dimension, Physical Review E, \newblock{\bf51}, R2705, 1995. 
\\Structure of avalanched and breakdown of simple scaling in the abelian sandpile model in one dimension, Physiacl Review E, \newblock{\bf52}, 4804, 1995. 
%\bibitem{maya}M. Paczuski and Stefan Boettcher, Universality in Sandpiles, Interface depinning, and Earthquake models, Physical Review Letters \newblock{\bf77}, \newblock 111 (1996).
%\bibitem{amaral}L. A. N. Amaral and K. Lauritsen, Self-organized criticality in a rice-pile model, Physical Review E \newblock{\bf54}, \newblock R4512 (1996).
%\bibitem{kim}K. Christensen, A. Corral, V. Frette, J. Feder, and Torstein Jossang, Tracer dispersion in a Self-Organized Critical system, Physical Review Letters \newblock{\bf77}, \newblock 107 (1996). 
%\bibitem{vespig}A. Vespignani, Stefan Zapperi, How self-organized criticality works: A unified mean-field picture, Physical Review E \newblock{\bf57}, \newblock 6345(1998).
%\bibitem{bak}P. Bak and K. Chen, The physics of fractals, Physica D \newblock{\bf38}, \newblock 5 (1989).
%\bibitem{chen}K. Chen, P. Bak and S. P. Obukhov, Self-organized criticality in a crack propagation model of earthquakes, Physical Review A \newblock{\bf43}, \newblock 625 (1991).
%\bibitem{sornette}A. Sornette and D. Sornette, Self-organised criticality of earthquakes, Europhysics Letters \newblock{\bf9}, \newblock 197 (1989).
\bibitem{feller} W. Feller, {\it An Introduction to Probability Theory and Its Applications} (John Wiley and Sons, New York, 3rd Edition, 1967, Vol.\newblock{\bf1}, p.\newblock 78).   

\end{thebibliography}

\end{document}